\documentclass[iop, apj, revtex4]{emulateapj}
\usepackage{natbib}
\usepackage{url}
\usepackage{amsmath}
\usepackage{appendix}
\usepackage{hyperref}

\newcommand{\kms}{km s$^{-1}$}

\begin{document}

\title{ALMA CO Clouds and Young Star Complexes in the Interacting Galaxies IC 2163 and NGC 2207}

\author{Debra Meloy Elmegreen\altaffilmark{1}, Bruce G. Elmegreen\altaffilmark{2},
Michele Kaufman\altaffilmark{3},  Elias Brinks\altaffilmark{4}, Curtis
Struck\altaffilmark{5}, Fr\'ed\'eric Bournaud\altaffilmark{6}, Kartik
Sheth\altaffilmark{7}, and Stephanie Juneau\altaffilmark{6}}

\altaffiltext{1}{Department of Physics \& Astronomy, Vassar College, Poughkeepsie,
NY 12604; elmegreen@vassar.edu}

\altaffiltext{2}{IBM Research Division, T.J. Watson Research Center, P.O. Box 218,
Yorktown Heights, NY 10598; bge@us.ibm.com}

\altaffiltext{3}{110 Westchester Rd, Newton, MA 02458, USA;
kaufmanrallis@icloud.com}

\altaffiltext{4}{{University of Hertfordshire, Centre for Astrophysics Research,
College Lane, Hatfield AL10~~9AB, United Kingdom; e.brinks@herts.ac.uk} }

\altaffiltext{5}{{Department of Physics \& Astronomy, Iowa State University, Ames,
IA 50011; struck@iastate.edu} }

\altaffiltext{6}{Laboratoire AIM-Paris-Saclay, CEA/DSM-CNRS-Universit\'e Paris
Diderot, Irfu/Service d'Astrophysique, CEA Saclay, Orme des Merisiers, F-91191 Gif
sur Yvette, France; frederic.bournaud@gmail.com; stephanie.juneau@cea.fr}

\altaffiltext{7}{NASA Headquarters, Washington, DC}

\begin{abstract}
ALMA observations of CO(1-0) emission in the interacting galaxies IC 2163 and NGC
2207 are used to determine the properties of molecular clouds and their association
with star-forming regions observed with the Hubble Space Telescope. Half of the CO
mass is in 249 clouds each more massive than $4.0\times10^5\;M_\odot$.  The
mass distribution functions for the CO clouds and star complexes in a
galactic-scale shock front in IC 2163 both have a slope on a log-log plot of -0.7,
similar to what is observed in Milky Way clouds. The molecular cloud mass function
is steeper in NGC 2207. The CO distribution in NGC 2207 also includes a nuclear
ring, a mini-bar, and a mini-starburst region that dominates the 24$\mu m$, radio,
and H$\alpha$ emission in both galaxies.  The ratio of the sum of
the masses of star complexes younger than 30 Myr to the associated molecular cloud
masses is $\sim4$\%. The maximum age of star complexes in the galactic-scale shock
front in IC 2163 is about 200 Myr, the same as the interaction time of the two
galaxies, suggesting the destruction of older complexes in the eyelids.

\keywords{ISM: molecules --- Galaxies: star formation}
\end{abstract}

\section{Introduction}
\label{intro}

The spiral galaxies IC 2163 and NGC 2207 are a well-studied pair undergoing a
grazing collision, previously observed in the X-ray, UV, optical, infrared, mm, and
cm wavelengths
\citep{kaufman12,elmegreen00,elmegreen01,elmegreen06,elmegreen95a,elmegreen95b,elmegreen16}
and modeled in simulations \citep{elmegreen95b,struck05}. The close encounter
produced in-plane tidal forces in IC 2163, resulting in a large shock producing a
cuspy-oval or ``eyelid'' structure at mid-radius and long tidal arms. The encounter
also produced forces nearly orthogonal to the plane of NGC 2207, resulting in a
warp. Details of the interaction are given in the cited papers.

Atacama Large Millimeter Array (ALMA) observations of $^{12}$CO (1-0) were
presented in two recent papers. Paper I \citep{elmegreen16} studied the
Kennicutt-Schmidt \citep{kennicutt12} relation for star formation and suggested
that some regions in NGC 2207 with high star formation rates ($>10^{-2}\;\rm
M_\odot$ pc$^{-2}$ Myr$^{-1}$) have a fraction of molecular to total gas of less
than 0.5 as a result of unusually high turbulent speeds. Paper II \citep{kaufman16}
showed $\sim100$ km s$^{-1}$ streaming motions in the eyelids that trace the pile-up
of molecular gas. Here in Section 2 we investigate the individual molecular clouds
seen with ALMA and measure their associated star complexes and OB associations.
Section 3 discusses Feature {\it i} and the nuclear region of NGC 2207. The
conclusions are in Section 4.

\section{Observations and Analysis of CO Clouds}
\subsection{Data}
IC 2163 and NGC 2207 are at a distance of 35 Mpc (NASA/IPAC Extragalactic Database,
NED). As described in Papers I and II, ALMA observations of the 2.6 mm line from
the $^{12}$CO (1-0) transition were made in 34 pointings, with a point spread
function of $2.00^{\prime\prime} \times 1.52^{\prime\prime}$ (HPBW), a channel
width of $10$ km s$^{-1}$, and an rms noise per channel of $3.7$ mJy beam$^{-1}$.
Further data reduction techniques are given in Papers I and II. The Hubble Space
Telescope (HST) WFPC2 data used in this paper are from our observations
\citep{elmegreen00,elmegreen01} taken with filters F336W (U), F439W (B), F555W (V),
and F814W (I). The IR data are from Spitzer IRAC observations at 8$\mu m$ and MIPS
observations at 24$\mu m$  \citep{elmegreen06}. Radio continuum observations at 6
cm observed with the Karl G. Jansky Very Large Array (VLA)\footnote{The National
Radio Astronomy Observatory is a facility of the National Science Foundation
operated under cooperative agreement by Associated Universities, Inc.}
 are described in \cite{kaufman12}.

\begin{figure*}\epsscale{1.15} \plotone{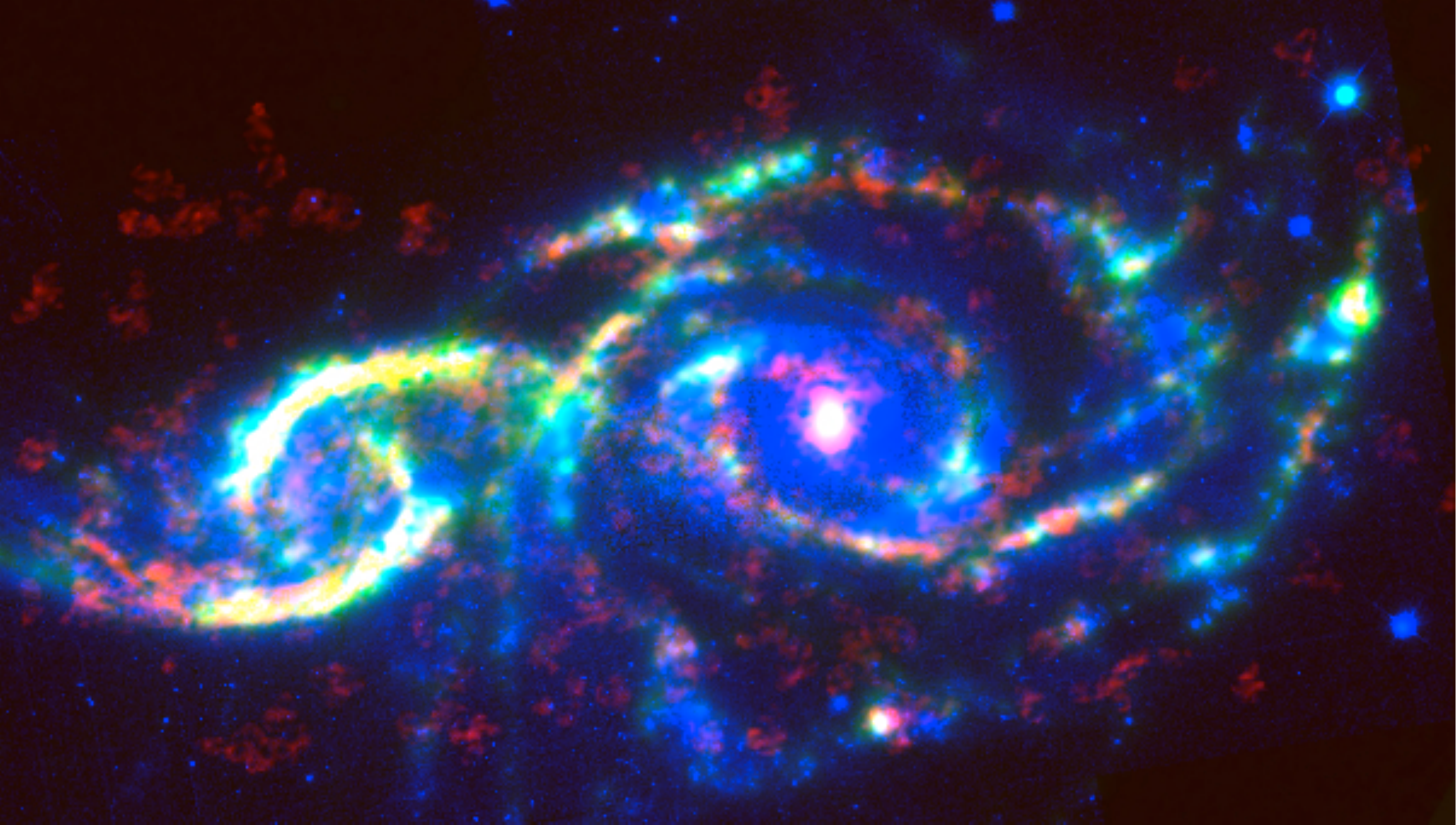}
\caption{Color composite showing HST WFPC2 F336W in blue, Spitzer IRAC channel 4
(8$\mu m$) in green, and ALMA $^{12}$CO (1-0) in red.  N is up. The ``eyelids"
of IC 2163 are the bright yellow arms on the East (left). Feature {\it i} in NGC
2207 is the bright yellow circular region in the tip of the Western (right) arm of
NGC 2207. }\label{I2163B8CO}\end{figure*}

Figure \ref{I2163B8CO} shows a composite color image with F336W in blue, IRAC
channel 4 (8$\mu m$) in green, and CO in red. (Figure 1 in Paper II showed the CO
emission overlaid on the F439W image.) The strongest CO emission is found in the
eyelids of IC 2163 and ``Feature {\it i}'' (Sect. \ref{featurei}) in the northwest
tip of NGC 2207; these features are bright in all passbands, as is the nuclear
region of NGC 2207. CO emission also appears in smaller clouds throughout the
galaxies, most prominently in the arms.

\subsection{ Identification of CO Clouds}

\begin{figure*}\epsscale{1.15} \plotone{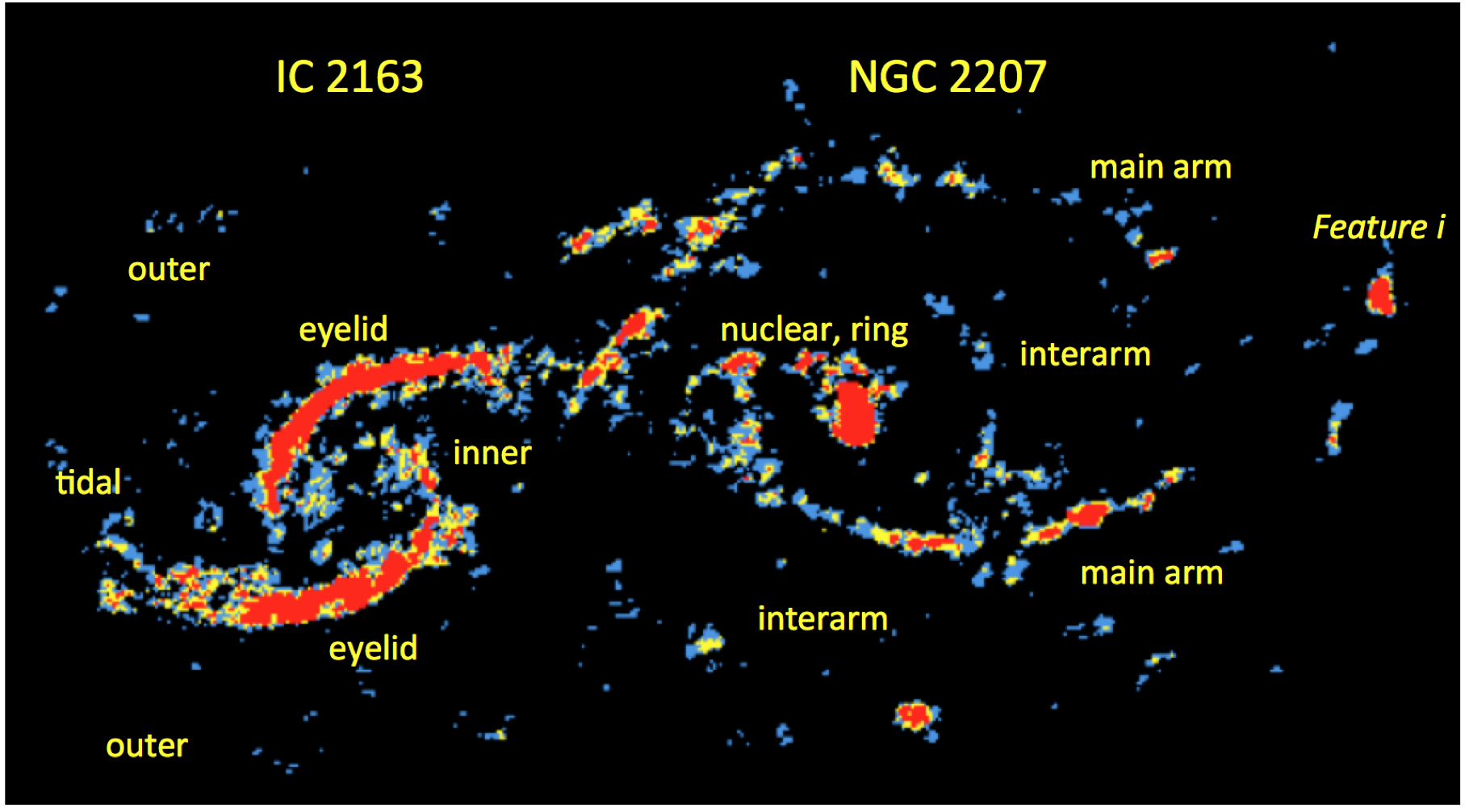}
\caption {This is a color-coded display of the CO emission
where the color represents the number of channels brighter than some noise level.
Black pixels in the figure are for positions with 0 or 1 velocity channels brighter
than $2\sigma$,  blue is for 2 channels brighter than $2\sigma$, yellow is for 3
channels brighter than $2\sigma$, and red is for 4 or more channels brighter than $2\sigma$. }
 \label{COannot}\end{figure*}

Paper I determined the total molecular mass detected by our
interferometric observations.  For IC 2163, the H$_2$ mass is $2.1\times10^9\; M_\odot$ and for
NGC 2207, it is $2.3\times10^9\;M_\odot$. Here, we focus on individual CO
clouds, identified using several steps involving either intensity or velocity data. After creating a blanking mask via the
technique of smooth-plus-velocity continuity (cf. Papers I and II) and applying it
to the original unsmoothed cube, we identified 327 discrete CO clouds by eye in the
resulting surface density image in DS9.  To be conservative in setting a limit for
discriminating between clouds and noise spikes, we then
considered in the following analysis only clouds that contained 2 or
more pixels (2 px = 1.0'' = 170 pc at the assumed distance of 35 Mpc) above a
CO intensity $I(CO)= 200$ Jy beam$^{-1}$ m s$^{-1}$, which corresponds to a
line-of-sight (L.O.S.) column density N(H$_2$) of 17.7 M$_{\odot}$ pc$^{-2}$. Many
of the small clouds north and south of the eyelids of IC 2163 and in the interarm regions of NGC
2207 are below the selected contour limit. There are 26 such faint
clouds, and they amount to 1.3\% of the summed cloud mass. Without these clouds,
there were 301 clouds remaining in the sample.

Next, we eliminated clouds that have at most one velocity channel above
$2\sigma$, where $\sigma$ is the rms noise (= 3.7 mJy beam$^{-1}$ = 0.11 K) in the
unsmoothed cube, and we eliminated additional clouds that are smaller
than our spatial resolution of 2.0''$\times$1.5''. The final number of clouds
 was then 249;  these are used in the following
analysis. They are the dense, large-scale, CO-emitting regions that
stand out amidst the more diffuse CO emission in the galaxies. Such regions are
sometimes referred to as Giant Molecular Associations (GMAs), but we call them
clouds here.

Figure \ref{COannot} is a color-coded display of the CO emission,
where the color represents the number of channels brighter than a given noise level.
Black pixels in the figure are for positions with 0 or 1 velocity channels brighter
than $2\sigma$, blue is for 2 channels brighter than $2\sigma$, yellow is for 3
channels brighter than $2\sigma$, and red is for 4 or more channels brighter than $2\sigma$.
All of the clouds discussed in this
paper are in these colored regions.

 In a separate search for clouds, we also ran the CO intensity
image through the source extraction program SExtractor \citep{bertin}. The
parameter DETECT\_MINAREA, which is the minimum number of pixels above threshold,
was set to 20; the analysis threshold, ANALYSIS\_THRESH, was set to $10\sigma_{\rm
SEX}$ for rms $\sigma_{\rm SEX}$ defined by SExtractor, and the deblending
threshold, DEBLEND\_THRESH, was set to 32. This procedure yielded 178 CO clouds.
With these parameters, SExtractor detected the well-separated CO clouds that we
identified by eye, but the segmentation maps from SExtractor put all of the CO in
the eyelids of IC 2163 into a single cloud even though discrete CO clouds can be
seen clearly. SExtractor also put all of the CO in the southern inner main arm of
NGC 2207 into a single cloud. Separation of these bright regions into clouds
required different thresholds. In what follows we use the visual identifications of
clouds.

\subsection{CO Masses}

 Photometry on the visually-identified CO clouds was done with the
Image Reduction and Analysis Facility (IRAF) task {\it imstat},  in which a
variable rectangle is drawn around each cloud in order to match the cloud size
based on isophotal limits; this task then determines a total intensity based on the
average intensity times the number of pixels. This method is analogous to aperture
photometry using a variable aperture, but is useful for outlining elongated cloud
regions with rectangles instead of circles.  Background was not subtracted
because the background was already blanked in this image as a result of applying a
mask to the CO cube. To convert the CO emission into molecular mass $M(H_2)$, we
use $X_{\rm CO}= 1.8 \pm 0.3 \times10^{20}$ $H_2$ cm$^{-2}$ (K \kms)$^{-1}$ from
\cite{dame01}.  All cloud masses given in this paper are $H_2$
masses and should be multiplied by 1.36 to get the total gas mass, including helium
and heavy elements.  The sum of the cloud $H_2$ masses is $ 7.4\times10^8\;M_\odot$
for IC 2163 and $9.1\times10^8\;M_\odot$ for NGC 2207. These are 35\% and 40\%,
respectively, of the total $H_2$ masses measured in Paper I.

At the ALMA resolution of $2.00^{\prime\prime}\times1.52^{\prime\prime}$, which is
$340\times260$ pc,  anything detected here as a CO cloud is a large
region of molecular material; $10^6\;M_\odot$ spread out over the beam size
corresponds to an average surface density of $11\;M_\odot$ pc$^{-2}$, which is only
about one-tenth the surface density of a typical Milky Way CO cloud.

\begin{figure}\epsscale{1.} \plotone{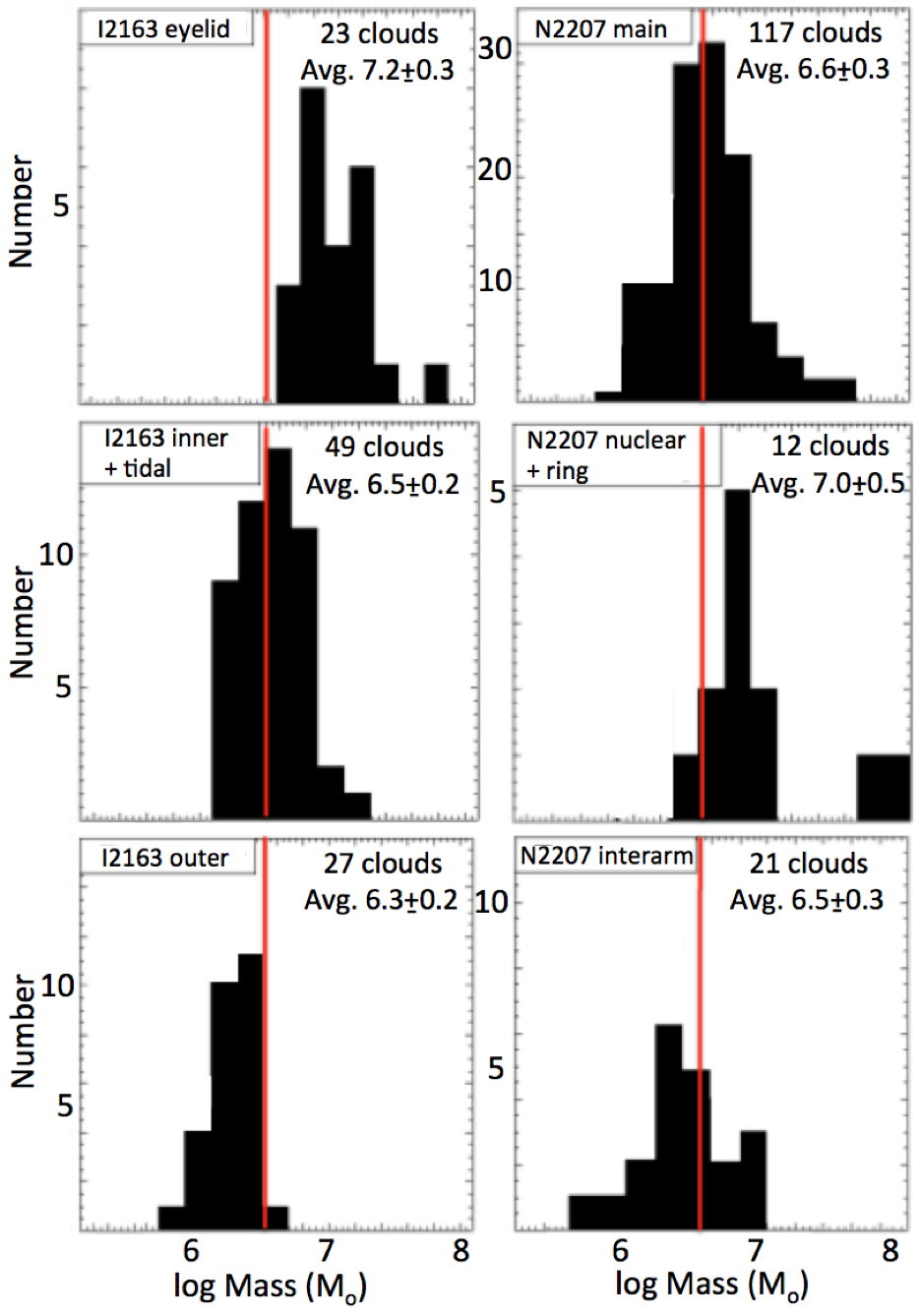}
\caption{Histogram of the logs of the cloud masses for different regions of IC 2163
(eyelids, inner plus tidal arms, outer regions) and NGC 2207 (main arms, inner ring,
inter arms). The red line is a fiducial marker for $\log_{10}(M/M_\odot)= 6.5$, which is the
average for the inner plus tidal arm clouds of IC 2163 and the interarm clouds of
NGC 2207. The number of clouds and their average masses in the logarithm are
indicated for each region.} \label{COhist}\end{figure}

For comparison purposes, the clouds are divided into six large regions. Three regions
are in IC 2163: a) the eyelids, b) the inner arms plus the tidal arms, and c) the outer
regions beyond the arms. In NGC 2207, the three regions are a)  the main spiral ams, b)
the nuclear region plus nuclear ring, and c) the interarm regions. Figure \ref{COhist}
shows histograms of the logs of the cloud masses for the six regions, binned in
$\log_{10}(M/M_\odot)$ units of 0.2. For reference, a red line with $\log_{10}(M/M_\odot)= 6.5$ is shown in
each plot; this is the average value for the inner and tidal arms of IC 2163 and the
main arms of NGC 2207. The minimum detected cloud mass was $\log_{10}(M/M_\odot)=5.6$. The most massive cloud in either galaxy
 is in the nuclear region of NGC 2207, with $\log_{10}(M/M_\odot)=8.0$. The eyelid regions of IC 2163 contain clouds with the
highest average, with $\log_{10}(M/M_\odot)=7.2$; all have $\log_{10}(M/M_\odot)$ greater than 6.6, and twelve have$\log_{10}(M/M_\odot)=7$ or greater.
 The tidal arms of IC 2163 have cloud masses similar to those in the main arms of
NGC 2207, and the clouds in the outer, non-arm regions of both galaxies are less
massive and also similar to each other in mass.

It is interesting to compare these masses with high resolution studies of other
interacting galaxies. \cite{koda} observed CO in M51 using CARMA and NRO with an
angular resolution corresponding to 160 pc. In their Figure 2, they show the distribution of giant molecular clouds
(GMCs) which they define as having $\log_{10}(M/M_\odot)$ of 5 to 6, and of giant molecular
associations (GMAs), with log masses of 7 to 8. The GMAs are located exclusively in the
spiral arms of M51, as they are in the eyelid regions of IC 2163 (and a few in the main
arms of NGC 2207) in this study. Their GMCs are also mostly in the M51 arms, but there
are many in the interarm regions, similar to the distribution of lower mass clouds in
IC 2163 and NGC 2207. They conclude that GMAs have lifetimes of a few 10 million years,
associated with the arm crossing times. \cite{wei} used the Submillimeter Array and the
Plateau de Bure Interferometer to observe CO(2-1) with an angular resolution of 160 pc
in the Antennae colliding galaxies, NGC 4038 and NGC 4039. They find $\log_{10}(M/M_\odot)$ ranging
from 5 to 8, with the most massive clouds in the regions of the most intense star
formation. 

\begin{figure}\epsscale{1.2} \plotone{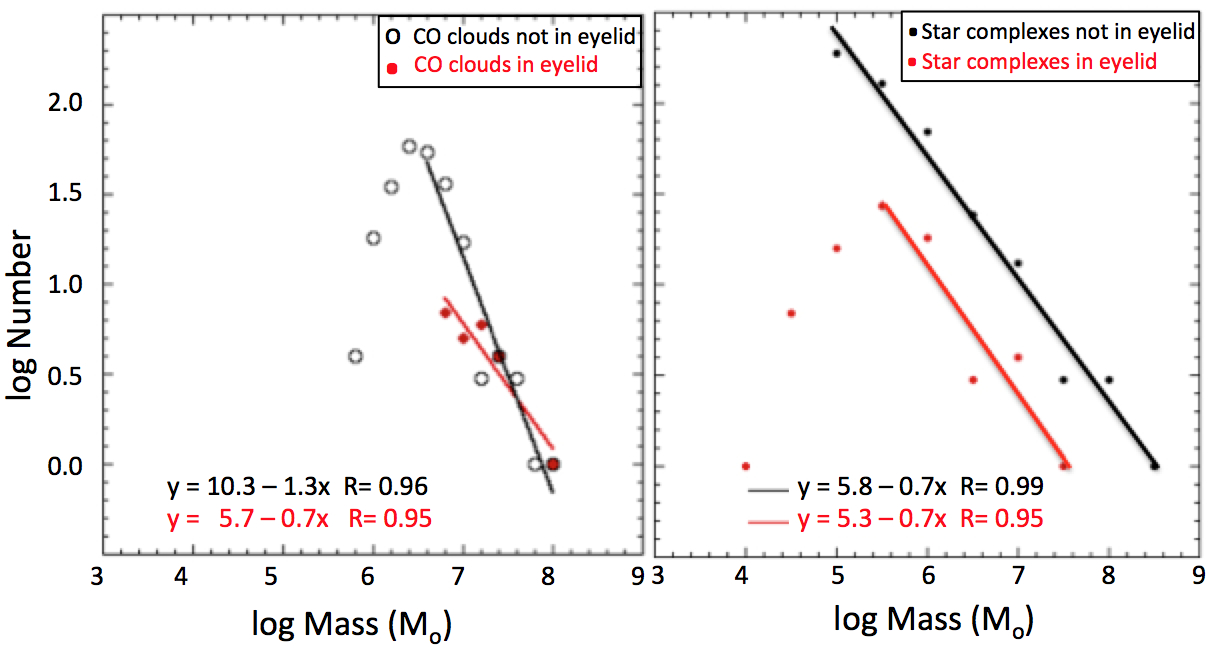}

\caption{(left) log(Number) vs.$\log_{10}(M/M_\odot)$ for all clouds in both IC 2163 and NGC 2207, binned in $\log_{10}(M/M_\odot)$
intervals of 0.2.  The red dots are clouds in the eyelids of IC 2163 only; the black dots
are the rest of the clouds in both galaxies. The power law slope is $-0.7\pm 0.1$ for the eyelid
clouds and $-1.3\pm 0.1$ for the rest of the clouds. (right) log(Number) vs. $\log_{10}(M/M_\odot)$ for
star complexes (black dots) in both galaxies except for the eyelids, and only in the eyelids
(red dots), binned in $\log_{10}(M/M_\odot)$  intervals of 0.5. The power law slope is the same for both,
$-0.7\pm 0.1$. } \label{allhist}\end{figure}

The combined cloud mass function is shown in the left-hand panel of Figure
\ref{allhist}, binned in $\log_{10}(M/M_\odot)$ intervals of 0.2. Clouds in the eyelids are shown as
red dots, while all others are shown as black circles. The linear fitting functions for
each sample are shown in the matching color. The eyelid cloud mass function has a slope
of $-0.7\pm 0.1$, while the rest of the clouds have a slope of $-1.3\pm 0.1$. For only
the clouds in the main arms of NGC 2207, shown in the upper right panel of Figure
\ref{COhist}, the slope is $-1.6\pm 0.1$. For reference, the mass function for clouds
identified by SExtractor has the same slope in the$\log_{10}(M/M_\odot)$ range of 6.6 to 7.6
($-1.3$ with a goodness of fit R=0.99) as the mass function for the visually identified
clouds in Figure \ref{allhist}.

The large slopes outside of the eyelid region are steep compared to the mass
function of inner Milky Way molecular clouds, where the power law slope for equal
intervals of $\log_{10}(M/M_\odot)$ is between $-0.4$ and $-0.6$ \citep{rosolowsky05}; they are more
comparable to those estimated for the outer Milky Way, which range between $-1$ and
$-1.6$ \citep{rosolowsky05}. In M33, the slope was measured to be $-1.85$ by
\cite{rosolowsky05}, $-1.1$ between 2 and 4 kpc by \cite{rosolowsky07}, and to steepen
from $-0.6\pm0.2$ in the inner 2 kpc to $-1.3\pm0.2$ beyond 2 kpc by \cite{gratier12}. In
the Antennae, \cite{wei} derived an equally steep power law slope of $\sim-1.4$ up to
$10^{6.5}\;M_\odot$. In M51, however, the slopes are more like they are in the Milky
Way: \cite{colombo} measured CO (1-0) cloud masses at 40 pc resolution and derived mass
spectrum slopes ($\gamma+1$ in their notation) of $-0.6$ to $-0.8$ up to
$10^7\;M_\odot$ in the ring and density wave arms. They did find, however, that the
slope steepness for higher mass clouds. Some of these differences may reflect different
cloud identification methods. \cite{miville17}, for example, got a slope of
$-3.0\pm0.1$ for equal intervals of $\log M$ at the high mass end of Milky Way CO
clouds.

\subsection{Mass spectrum variations}

The reason for these variations in the CO cloud mass spectrum slope are unknown. The
power spectrum of cloud structure seems to be a nearly universal power law for a range
of scales up to the thickness of a galaxy \citep{stan99,stan00, elmegreen01, block10,
combes12,zhang12}. The mass function for clouds should follow from the power spectrum
slope of cloud structure \citep{shad11}. But this is the mass spectrum for total cloud
structure and not just the mass of the CO-emitting part. If the fraction of the cloud
mass in the form of CO decreases with increasing cloud mass, then the CO cloud mass
spectrum will be steeper than the total cloud mass spectrum. For example, Paper I
suggested that some star-forming regions with high SFR in NGC 2207 are HI dominated
with relatively small CO cores. The largest clouds in the non-eyelid regions would then
have a lower molecular fraction than the clouds in the eyelids. This could cause the
molecular mass spectrum to drop more quickly than the total cloud mass spectrum in the
non-eyelid regions.

A specific example where the molecular mass fraction per cloud decreases with
increasing cloud mass is if the cloud edge always has a density too low for CO
formation or excitation, and the density gradient inside the cloud gets steeper with
increasing total mass.  Then the CO core of the cloud would shrink relative to the
whole cloud as the total mass increased. This changing density gradient is reasonable
because the virial parameter $\alpha$ decreases with increasing mass in most CO surveys
\citep[e.g.,][]{miville17}, and lower virial parameter corresponds to stronger
self-gravity and a more concentrated cloud core.

To quantify this model, consider that the mass of the CO part of a cloud, $M_{\rm CO}$ (which in the previous section we 
converted to $M(H_2)$)
increases with the total cloud mass, $M_{\rm tot}$, as $M_{\rm CO}\propto M_{\rm
tot}^\zeta$ for $\zeta<1$. Suppose also that the total mass spectrum of interstellar
structure is $dN(M_{\rm tot})/dM_{\rm tot}\propto M_{\rm tot}^{-\beta}$. Then the mass
spectrum of the CO parts of the clouds may be derived from the one-to-one relation
between the two, $N(M_{\rm CO})dM_{\rm CO}=N(M_{\rm tot})dM_{\rm tot}$, from which we
obtain $dN(M_{\rm CO})/dM_{\rm CO}\propto M_{\rm CO}^p$ for $p=-(\beta+\zeta-1)/\zeta$.
If $\beta$ is in the likely range from 1.5 to 2, then $p$ is more negative than
$\beta$. That means that the mass spectrum of CO-identified ``clouds'' is steeper than
the mass spectrum of the physical cloud structure. The slope of $p\sim-2.3$ for CO
clouds derived above ($-1.3$ for equal intervals of $\log M$) requires $\zeta\sim0.4$
to $0.8$ for these $\beta$, respectively.

Another consideration is that for typical interstellar clouds, there is an inverse
relation between average cloud density and mass \citep{larson81}.  If there is a
corresponding decrease in molecular fraction at lower average cloud density, then this
CO mass fraction decreases with increasing total mass. For example, \cite{miville17}
derived a fractal mass-radius relation $M_{\rm CO}=36.7R_{\rm CO}^{2.2}$ for CO-cloud
radius $R_{\rm CO}$ in pc. This fractal power is also about the same as that determined
by other studies, even for atomic and dust masses \citep{sanchez05}. The
\cite{miville17} relation gives an average $H_2$ molecular density for the CO-part of
the cloud $n=620R^{-0.8}$ cm$^{-3}$, which is $n=2300M_{\rm CO}^{-0.36}$ cm$^{-3}$
(assuming a mean molecular weight of $4\times10^{-24}$ grams). Observations that trace
only molecules in an interstellar medium that is partially molecular should show a
rapid drop in the molecular cloud mass function as the average cloud density drops and
the clouds turn atomic at higher masses.  This explanation would apply to the outer
regions of galaxies mentioned above where the steep mass function slopes were observed.
It would not apply to the inner part of the Milky Way or M51 or the IC 2163 eyelids, where the slope is
shallower, because the clouds in these regions are highly molecular.

\subsection{Star-forming Complexes and CO Clouds}
In order to compare the locations and masses of the optical star-forming regions
with the locations and masses of the CO clouds, we ran SExtractor on the F439W
images for each galaxy to identify the most prominent star-forming regions. These
positions were then used to extract the corresponding sources in the F336W, F555W,
and F814W images.  Foreground stars were removed by using the task {\it imexamine}
to eliminate point sources with Gaussian profiles. There are 733 identified
star-forming regions. One pixel in the HST images is
$0.1^{\prime\prime}$, which corresponds to 16.9 pc. The selected star-forming
regions were restricted to 10 contiguous pixels greater than the
$10\sigma$ threshold, so the lower limit to the sizes of the regions is about 50
pc. Population synthesis fits to determine ages and masses based on the photometric
magnitudes and colors for solar abundance were performed using
the methods described in \cite{elm12}, with models from \cite{bruzual} including
extinction from \cite{calz00} and \cite{leith02}. The star
formation history for each region was assumed to be a decaying exponential with a
decay time of 10 Myr.

The mass distribution function for the star complexes is shown in Figure
\ref{allhist} (right), with log(Number) versus $\log_{10}(M/M_\odot)$ for the eyelid region
complexes (red dots) and the rest of the complexes (black dots), binned in log mass
intervals of 0.5. The fits to the high mass end are indicated by red and black
lines, respectively. The slopes are the same, $-0.7\pm 0.1$, even though the star
formation activity differs in these two types of regions. This slope is also the
same as for the mass distribution function of the CO clouds in the eyelids (Figure
\ref{allhist}, left).

For the Antennae galaxies, \cite{wei} compare the cloud mass function with the
super star cluster mass function, and conclude that the slopes are similar to each
other (-1.4 and -2, respectively). Thus, in the Antennae and in IC 2163 and NGC 2207,
slopes for mass functions in star forming regions are similar to the clouds in which
they formed.

\begin{figure}\epsscale{1.} \plotone{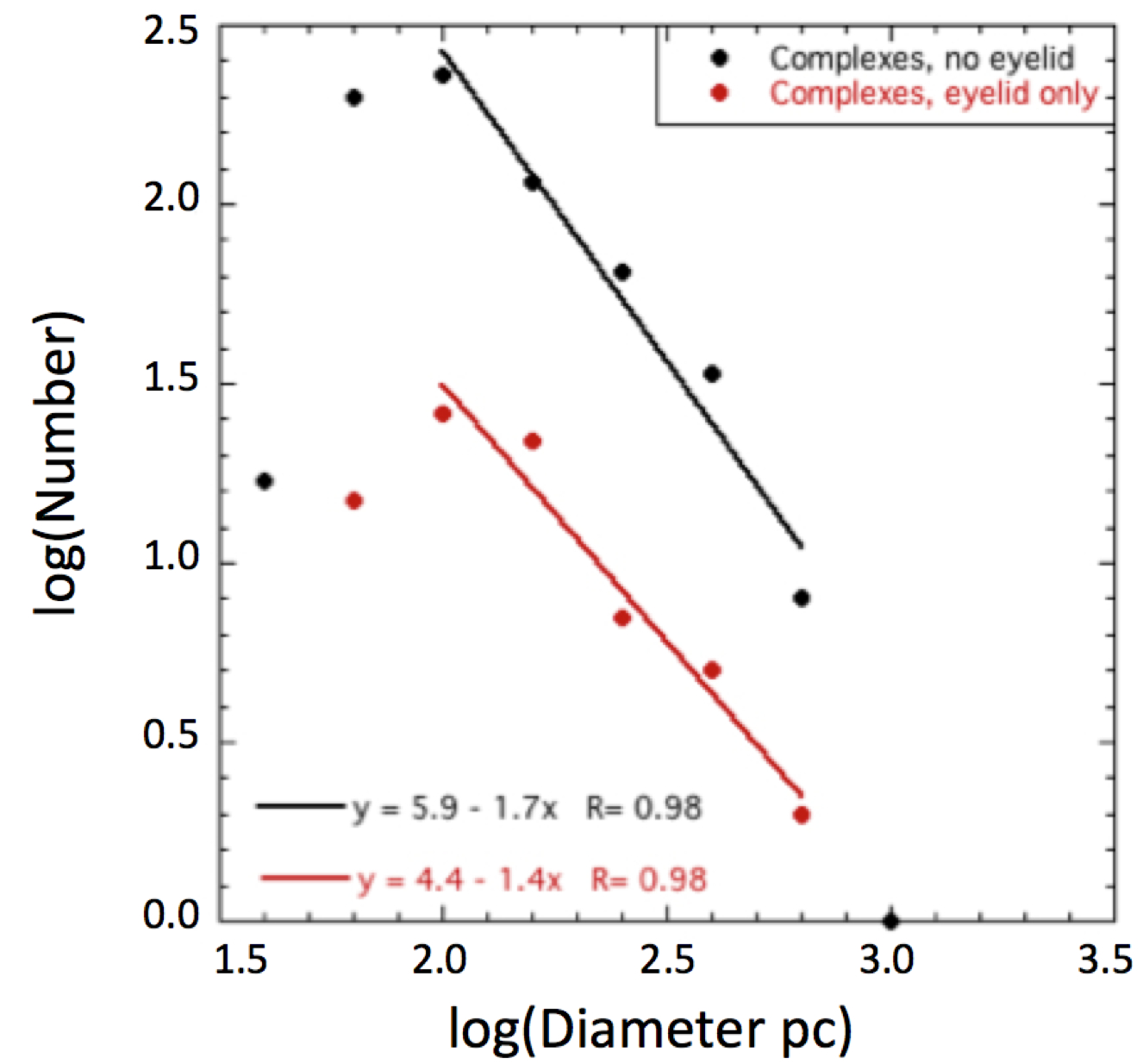}
\caption{Distribution of log(Number) versus log(Diameter) of SExtractor-fitted
star-forming complexes for the eyelid region (red) and the other regions combined
(black). Size is determined from the square root of the number of HST pixels. The
slopes from log(Diameter)=2.0 to 2.8 are indicated for the two samples. The
distributions turn over at the low size end, and there is only 1 complex at
log(3.0).} \label{starhist}\end{figure}

The size distributions of the SExtractor-fitted sources are shown in Figure
\ref{starhist}, binned in log(Diameter) intervals of 0.5.  Most fitted sources have
a diameter of about 100 pc, which means they are several pixels square in the
images. These are not star clusters, which would have a size of $\sim5$ pc or less,
but OB associations or giant star complexes \citep{efremov95}. Unlike star
complexes of this size in the Milky Way, which tend to be $\sim30-50$ Myr old, some
of the regions with the same size in IC 2163 and NGC 2207 are very young, with an
age of $\sim1$ Myr. Nevertheless, we refer to them as star complexes in what
follows. The size distribution in the eyelids has a slope of $-1.4$, compared with
a slope of $-1.7$ in the other regions of the galaxies, in the log(Diameter) range
of 2.0 to 2.8. The distributions turn over at the low size end from incompleteness;
there is only 1 complex in the 1 kpc size bin.

\begin{figure}\epsscale{1.15} \plotone{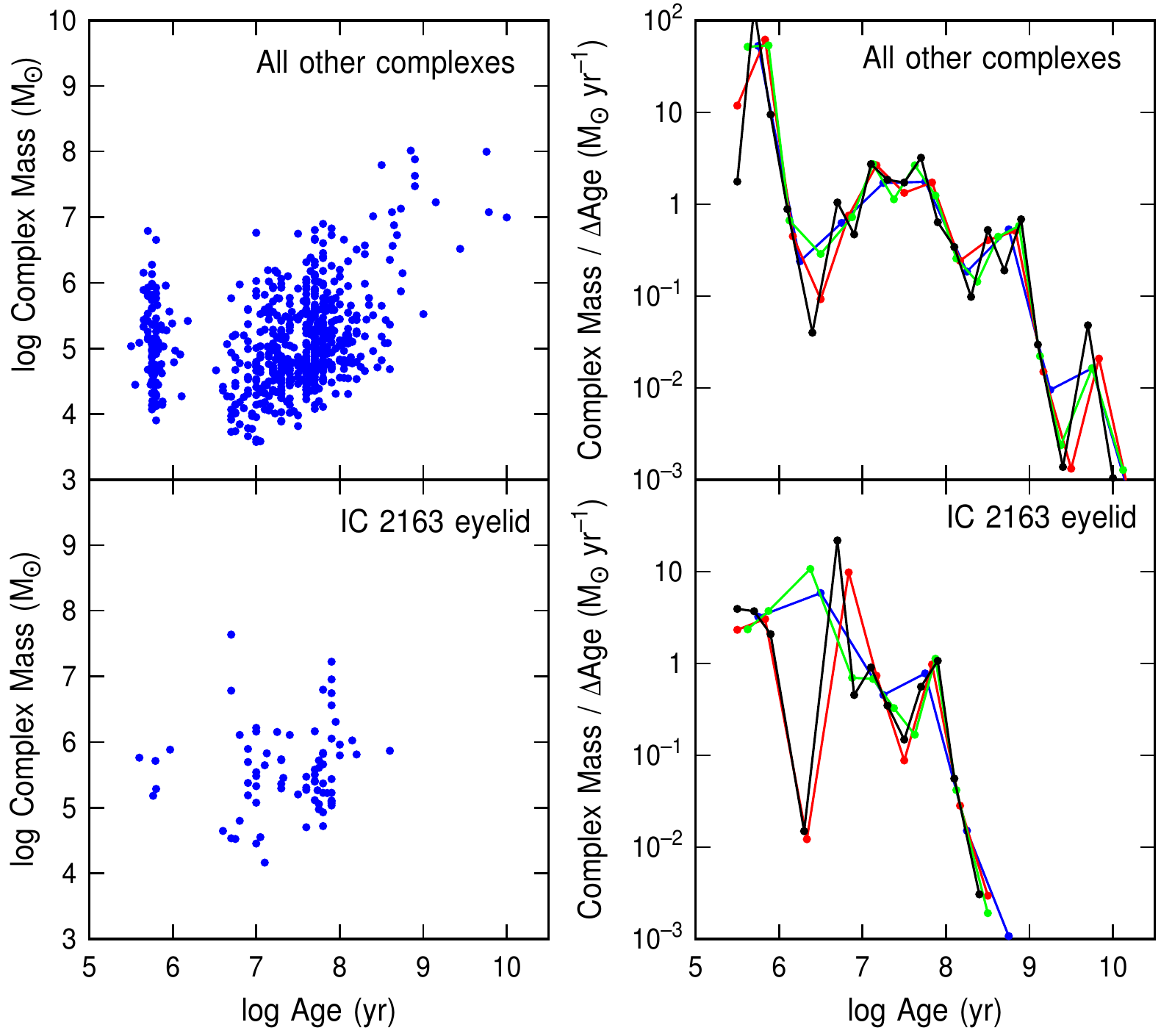}
\caption{(left) $\log_{10}(M/M_\odot)$ versus log(Age) for all star complexes outside the eyelids of IC 2163 and in NGC 2207 (top) and
inside the eyelids of IC 2163 (bottom). The models are not accurate for ages below $3\times 10^6$ yr, so the gap between log(Age) = 6 and 7 is an artifact.
(right) Summed complex mass per unit log(Age) interval versus log(Age)
for all star complexes except the eyelid regions (top) and only the eyelid regions (bottom). Different curves use
different bins of log(Age), noted in the text.}
 \label{starsmassage}\end{figure}

\begin{figure*}\epsscale{1.1} \plotone{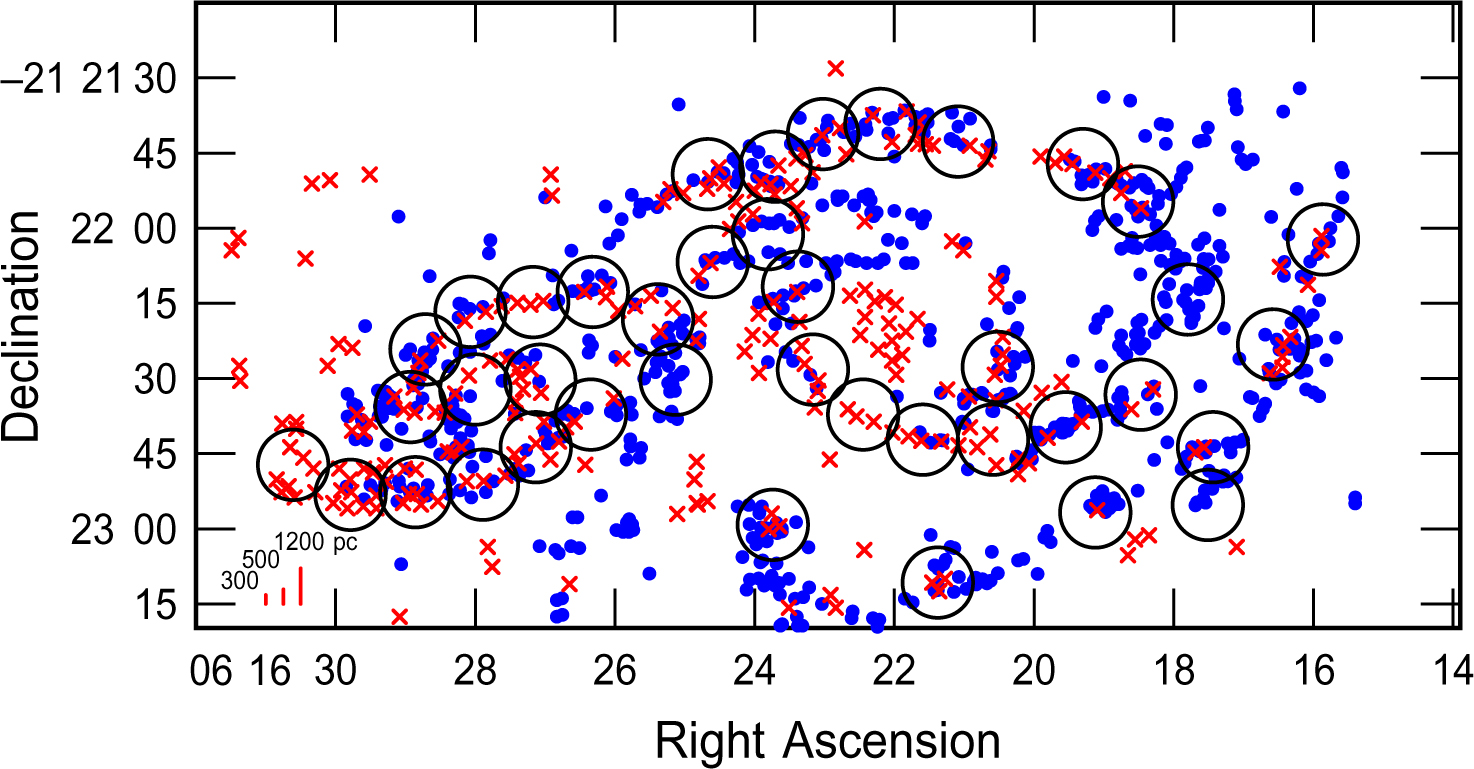}
\centering \caption{Comparison of CO and star complexes: red crosses represent CO
clouds from Figure \ref{COannot}, blue dots are star complexes, and black circles
are the 2400 pc diameter SF apertures with prominent 24$\mu m$ emission and/or
H$\alpha$ emission. The lines in the lower left show the scale for 300 pc, 500 pc,
and 1200 pc radii used in the text to determine which star complexes are near CO
clouds.} \label{COdot}\end{figure*}

Figure \ref{starsmassage} (left) shows a log-log plot of star complex photometric
mass as a function of photometric age for all complexes except those in the eyelids
(top) and for only those in the eyelids (bottom). The models are not
accurate for ages below $3\times10^6$ yrs, so these complexes will be ignored.
There is the usual trend of increasing mass with age. The lower mass
limit for this trend is because of star complex fading with time. The upper mass limit
is the size-of-sample effect, where the mass of the most massive complex increases
with the number of complexes, and the number of complexes per unit $\log$ time
increases with time when the formation rate is constant \citep{hunter03,elm10}.

Figure \ref{starsmassage} suggests there may be a cut-off in age for eyelid
complexes compared with the rest of the complexes. To determine whether this
cut-off is statistically significant, we consider the number of regions in various
age intervals. There is only 1 out of 77 star complexes in the eyelids that is older
than $2\times10^8$ yr, compared with 34 out of 656 at least this age in the other
parts of the galaxies.

The difference in complex ages for
the eyelids compared with elsewhere is shown in the right-hand side of Figure
\ref{starsmassage}, where the number of complexes in logarithmic age intervals of
various lengths, divided by the age intervals (thereby giving rates) are plotted
versus the logarithm of age for the eyelid and non-eyelid regions. The variation
among the different curves is a result of occasional gaps in the age distribution.
In order to determine the number per unit age interval even when there are no star
complexes in a particular interval, we include two or three nearby age intervals
until there are some complexes, and then divide that number of complexes by the
total included age interval. The statistical errors are relatively small for most
of the age intervals if we use the inverse square root of the number of complexes
as a measure of this error. However, in the ages where the rate drops below
$\sim0.1$, the error is of order unity, corresponding to only a few complexes per
age interval, as seen also in the left-hand panels. Thus, we estimate that the
falloff in the eyelid complex formation rate at $>2\times10^8$ yrs is a $4\sigma$
result, based on $5.2\pm0.9\%$ of all 656 complexes that old in non-eyelid regions
and 1 complex that old in the eyelids ($1.3\pm1.3\%$ of 77 eyelid complexes). To
make the eyelid and non-eyelid rate equal, we would need 3 more star complexes in
the eyelids with ages larger than $2\times10^8$ yrs.

The relatively small number of eyelid complexes older than $\sim2\times10^8$ yrs is
 interesting because our models \citep{struck05} suggest that
perigalacticon occurred $\sim2.4\times 10^8$ years ago. This then is compatible
with the encounter having destroyed the older eyelid complexes.

\subsection{Correlations between CO clouds and Star-forming regions}

The positions of the star complexes and clouds were cross-correlated to search for
star-forming regions within 300 pc and 500 pc of the CO clouds.  Figure \ref{COdot}
shows star-forming regions as blue dots and CO clouds as red crosses. Lines in the
lower left have lengths of 300, 500, and 1200 pc. Black open circles have 2400 pc
diameters and are used below to study the large-scale summed emission.

\begin{figure}\epsscale{1.15} \plotone{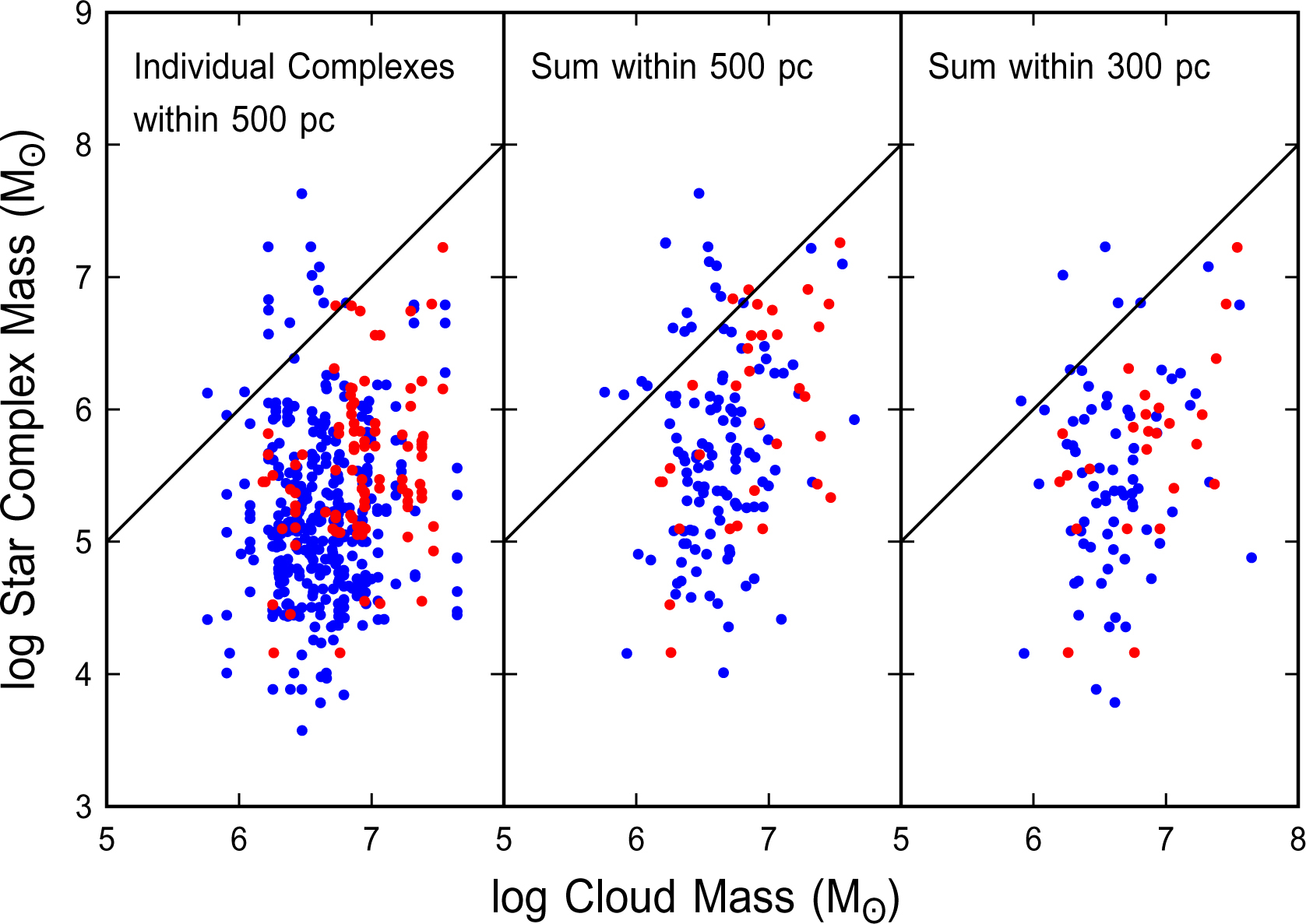}
\caption{Masses or summed masses of star complexes compared with nearby CO cloud masses. Red
dots are for star complexes in the eyelids, and blue dots are for other complexes.
(left) Individual complex masses for all complexes within 500 pc of a CO cloud
versus cloud mass. (middle) Sum of the masses of complexes within 500 pc of a CO
cloud versus cloud mass; (right) Sum of complexes within 300 pc of a CO cloud. The
lines represent a fiducial power law slope of unity with a one-to-one correspondence.} \label{COdotmass}\end{figure}

The cross-correlations are used to estimate the average star complex masses that
are associated with their adjacent cloud masses. Figure \ref{COdotmass} shows a
log-log plot of the star complex masses versus the cloud masses. On the left are
the CO cloud masses on the abscissa and the individual star complex masses within
500 pc on the ordinate. The red points correspond to regions in the IC2163 eyelids.
The average value of the difference between the log of the complex mass and the log
of the cloud mass is $-1.41\pm0.70$, which corresponds to an average ratio of
complex mass to cloud mass equal to $\sim3.9$\%. For the eyelid regions, it is
about the same, 4.3\%. In the middle panel are the sums of the masses of
star-forming regions within 500 pc of a particular CO cloud versus that cloud mass;
in the right panel are the sums of the masses of star-forming regions within 300 pc
of a CO cloud versus that cloud mass. Red points again represent the eyelids. The
lines are fiducial markers showing equality between the complex and cloud masses.
The average differences between the log of the summed complex masses and the log of
the cloud mass, within these two distances, are $-0.91\pm0.74$ and $-1.13\pm0.69$,
respectively; error bars are deviations around the mean. These correspond to ratios
of summed star complex mass to CO cloud mass of $\sim12$\% and $\sim7.4$\% for the
two neighborhoods.
For the eyelid regions, the ratios of complex/cloud mass are the same within the
errors, 12\% and 7.0\%.

These mass ratios are not the same as star formation efficiencies in
any one cloud because we do not know which complex formed in which cloud and
because some of the complexes are fairly old, $\sim100$ Myr. If we limit the
complex age to be less than 30 Myr and the distance to a cloud to be less than 300
pc, then the ratio of the summed complex mass to the cloud mass is 4.4\% (4.9\% for
IC 2163 complexes). These smaller numbers are more typical for star complexes.  The
higher efficiencies for larger regions may be explained if clouds are younger than
the complexes, which span a range of ages, and the complexes build up over time in
a sequence of different clouds. It is also possible that the molecular gas was
driven out of the regions in which the star complexes formed, or that star complex
formation occurred in denser molecular clouds of smaller mass.

\begin{figure}\epsscale{1.15} \plotone{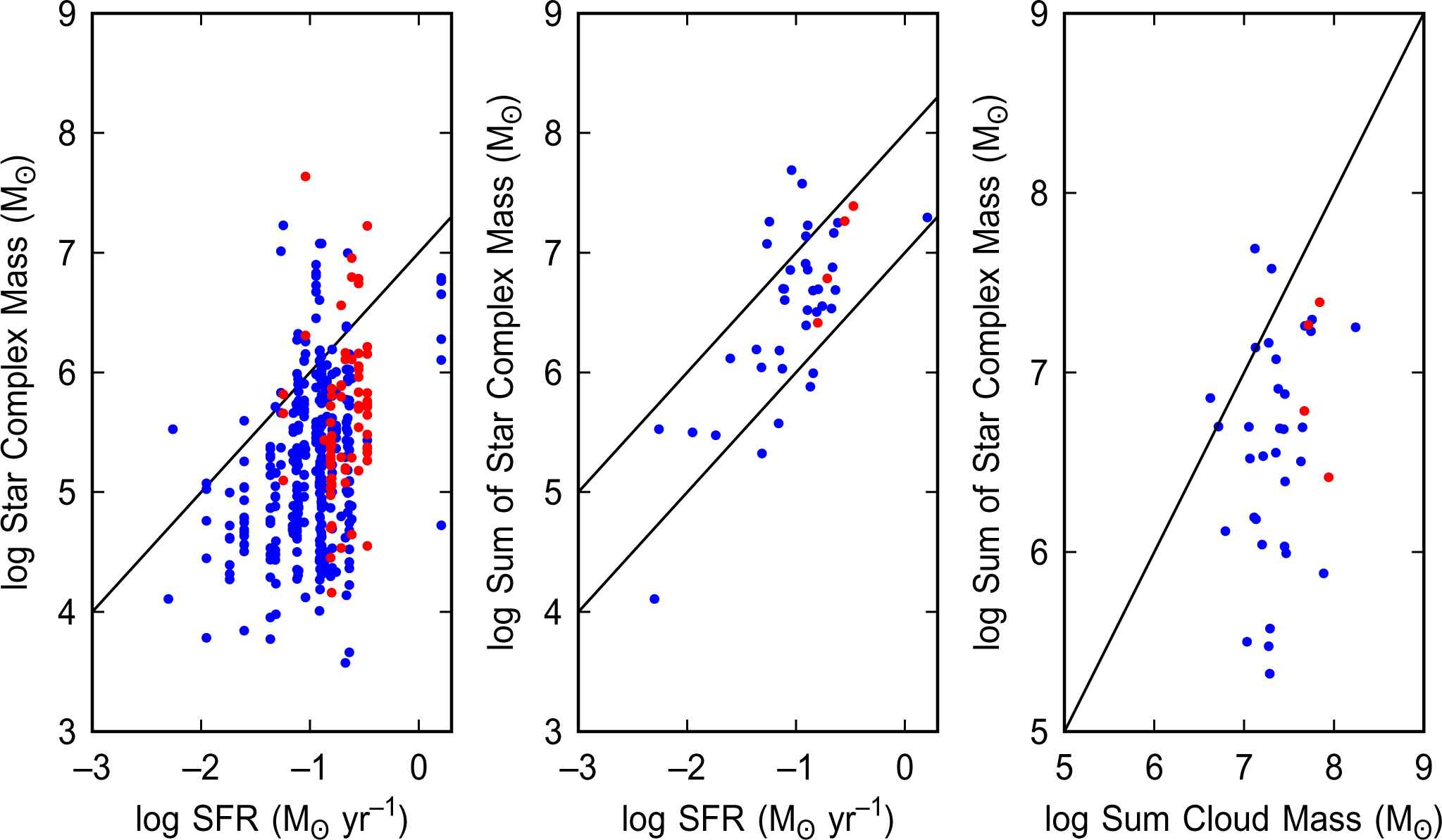}
\caption{Several properties are shown for optical star complexes and CO clouds within the 2400 pc
diameter apertures, as described in Paper I. In all panels, blue
dots represent star complexes outside of the eyelids, and red dots represent star
complexes in the eyelids of IC 2163.
The left panel shows star complexes
mass versus the SFR.  The middle panel shows the sum of the complex masses versus
the SFR. The right panel is the sum of the
complex masses versus the sum of the cloud masses within each SF aperture from
Paper I, and is equivalent to Figure \ref{COdotmass} on larger scales.
The lines have unit slope.}
\label{SFRcor}\end{figure}

Another way to examine the relationship between star-forming regions and CO clouds
is with fixed sizes centered on the large-scale star-forming regions. In Paper I we
defined regions enclosing prominent 24$\mu m$ and/or H$\alpha$ emission, and
measured the HI and H$_2$ masses and star formation rates at these positions. A
2400 pc diameter was chosen to match the resolution of our HI data. Here we
determine relative star complex masses on these scales by summing the complex
masses and the CO cloud masses within these regions, which we refer to as the ``SF
apertures''. These apertures are shown by large circles in Figure \ref{COdot}.

Most of the SF apertures contain both CO clouds and star complexes. There are 4
exceptions: SF apertures A1, A16, A26, and A39 (see Figure 1 in Paper I for
labels). A1 (on the eastern tidal arm of IC 2163) contains CO clouds and diffuse
star formation but no star complexes, and A26 (on the inner southern arm of NGC
2207 contains CO clouds, 8$\mu m$, 24$\mu m$, and diffuse H$\alpha$ emission, but
no star complexes.  A16 (between the two galaxies) and A39 (on the western outer
arm of NGC 2207) contain star complexes but no CO clouds. The remaining SF
apertures contain several CO clouds and star-forming complexes.

Comparisons of CO clouds and star-forming complexes within the SF
apertures are shown in Figure \ref{SFRcor}. The left-hand panel shows each star
complex mass versus the total star formation rate in the 2400 pc diameter aperture.
The masses of the largest star complexes increase with the star formation rate. The
line corresponds to the stellar mass that would result from the SFR on the abscissa
in $10^7$ years. Most complex masses are less than that, indicating that star
formation inside the regions is divided among several complexes which add together
to give the total rate. Some of the star formation in each SF aperture has probably
taken place outside the identified complexes too. Some complex masses greatly
exceed the expected value after $10^7$ yrs, suggesting a longer duration for the
accumulation of young stars.

The middle panel shows the sum of the complex masses within the 2400 pc diameter
aperture versus the star formation rate.  The lines represents the mass that would
be made by star formation in $10^7$ and $10^8$ years. This is a reasonable bracket
to the summed complex mass in each 2400 pc region, suggesting a duration of star
formation in about this range for the SF apertures.

The right-hand panel of Figure \ref{SFRcor} is the sum of the star complex masses
versus the sum of the cloud masses within each SF aperture, with the line showing a
fiducial slope of unity.  This plot is equivalent to Figure \ref{COdotmass}, but
now on a scale of 2400 pc diameter. The average difference between the logarithm of
the summed complex mass and the log of the summed cloud mass is $-0.78\pm0.62$,
which corresponds to a ratio of star complex mass to CO cloud mass equal to $17$\%.
With only star complexes less than 30 Myr, the mass fraction is 3.9\%.

These results for the ratios of young stellar mass to CO cloud mass in various
regions around the clouds suggest efficiencies of star formation over 30 Myr periods
that are between 3\% and 5\%.  This is normal for OB associations in nearby galaxies
\citep{molinari14}.  The ratios of stellar to molecular masses are larger when we
consider all of the stellar complexes near each molecular cloud, suggesting that
several generations of OB associations may form before the clouds are destroyed.

\subsection{Extinction}

\begin{figure}\epsscale{0.9} \plotone{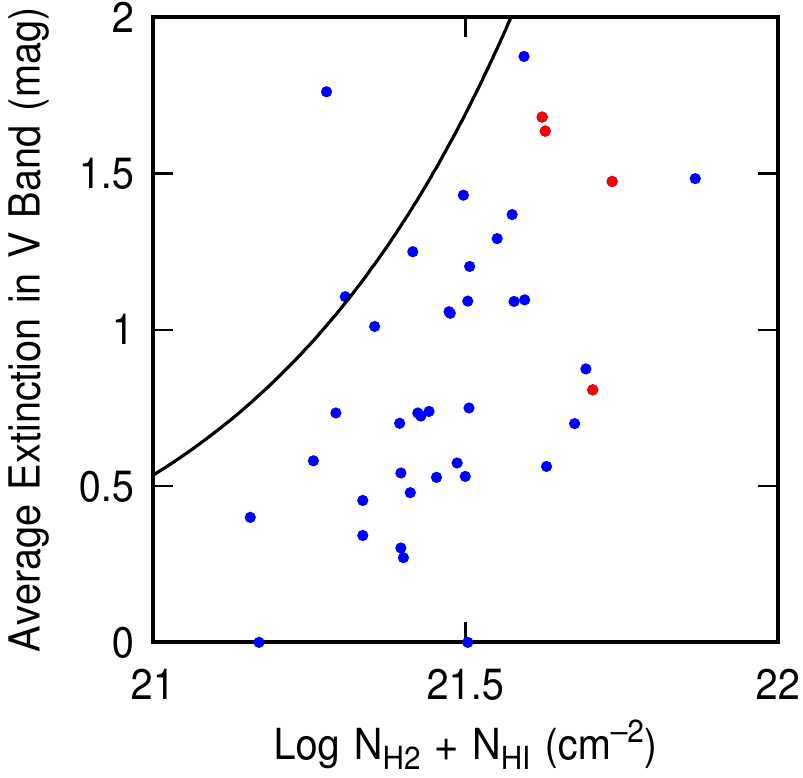}
\caption{The average extinction in V-band for the star complexes within a given SF aperture
is plotted versus the log of the average atomic + molecular column density for the
clouds within the aperture. The line shows the conversion between column density
and extinction based on $N(H)=5.8\times 10^{21} E_{B-V}$ \citep{bohlin78} and the
ratio of total to selective extinction $R_V=3.1$ \citep{draine03}.}
\label{extincvsdens}\end{figure}

Figure \ref{extincvsdens} shows the average V-band extinction in magnitudes for all
of the star complexes within a given SF aperture as measured from the
photometric fits versus the sum of the average line-of-sight atomic and molecular
column densities for the CO clouds in the aperture. The red dots represent the star
complexes in the eyelids, while the blue dots are for complexes not in the eyelids.
The line shows the conversion between column density and extinction assuming one magnitude of extinction at V-band corresponds to
$1.87\times10^{21}$ cm$^{-2}$ column density of hydrogen
\citep{bohlin78,draine03}.  The 14'' diameter apertures used to measure the
line-of-sight column densities average out local variations that could be present
in the immediate vicinity of the star complexes. On average, the extinction deduced
from the neutral hydrogen column densities is about twice the photometrically
determined value of A$_V$ of the star complexes. This is consistent with having the
star complexes in the midplane and the gas distributed symmetrically about the
midplane.The high point of extinction to the upper left of the curve is the central region of IC
2163.

\section{Other High Molecular Column Density Regions}
\subsection{Star Complexes in Feature {\it i}}
\label{featurei}

\begin{figure}\epsscale{1.2} \plotone{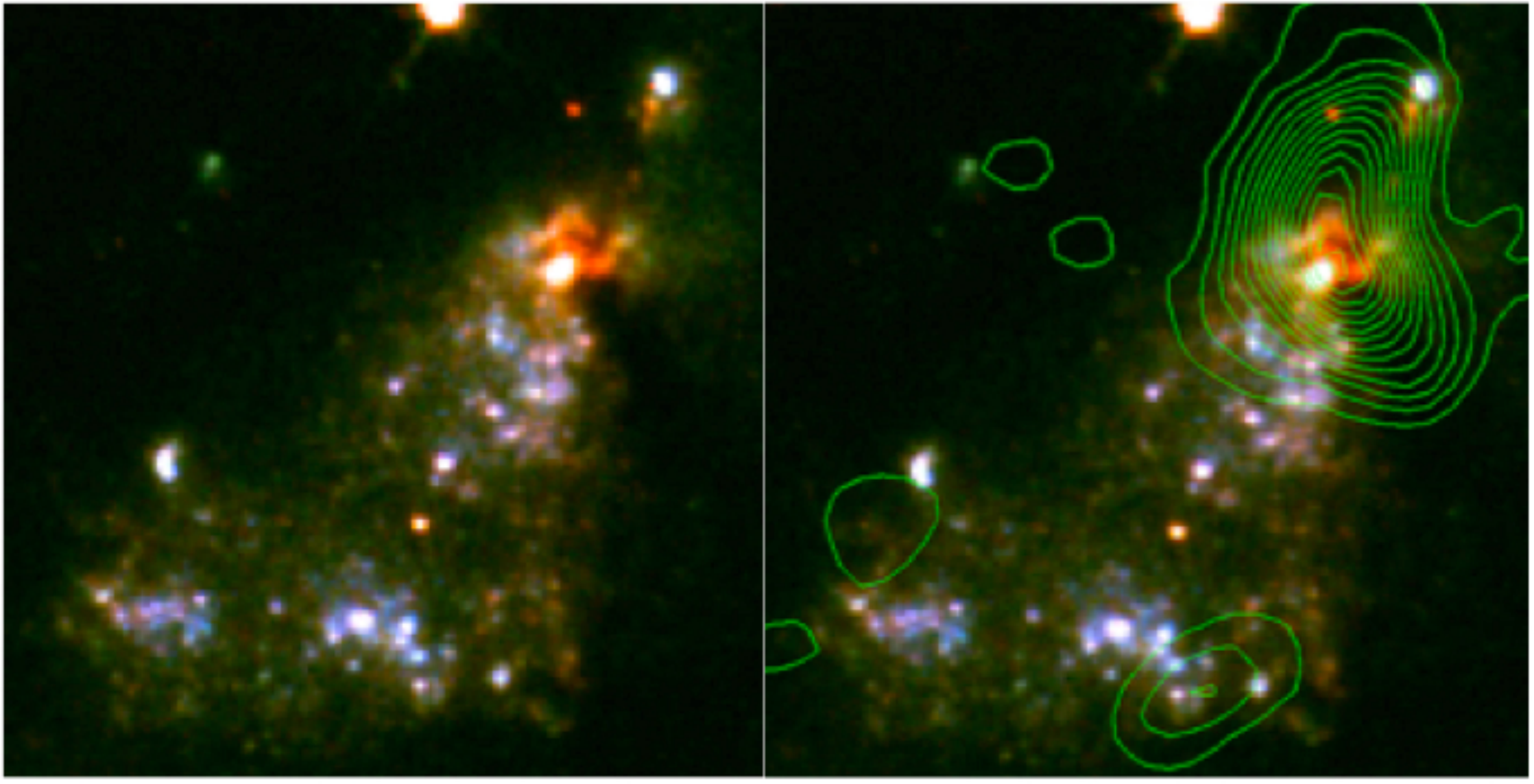}
\caption{(left) Feature {\it i} in the northwest arm of NGC 2207, shown in HST WFPC2 F439W,
F555W, F814W filters. (right) CO contours in green overlaid on the HST color image.
Contour flux values are in linear steps of 66 (Jy beam$^{-1}$) (m s$^{-1}$), or 5.8
M$_\odot$ pc$^{-2}$. } \label{feati}\end{figure}

\begin{figure}\epsscale{1.15} \plotone{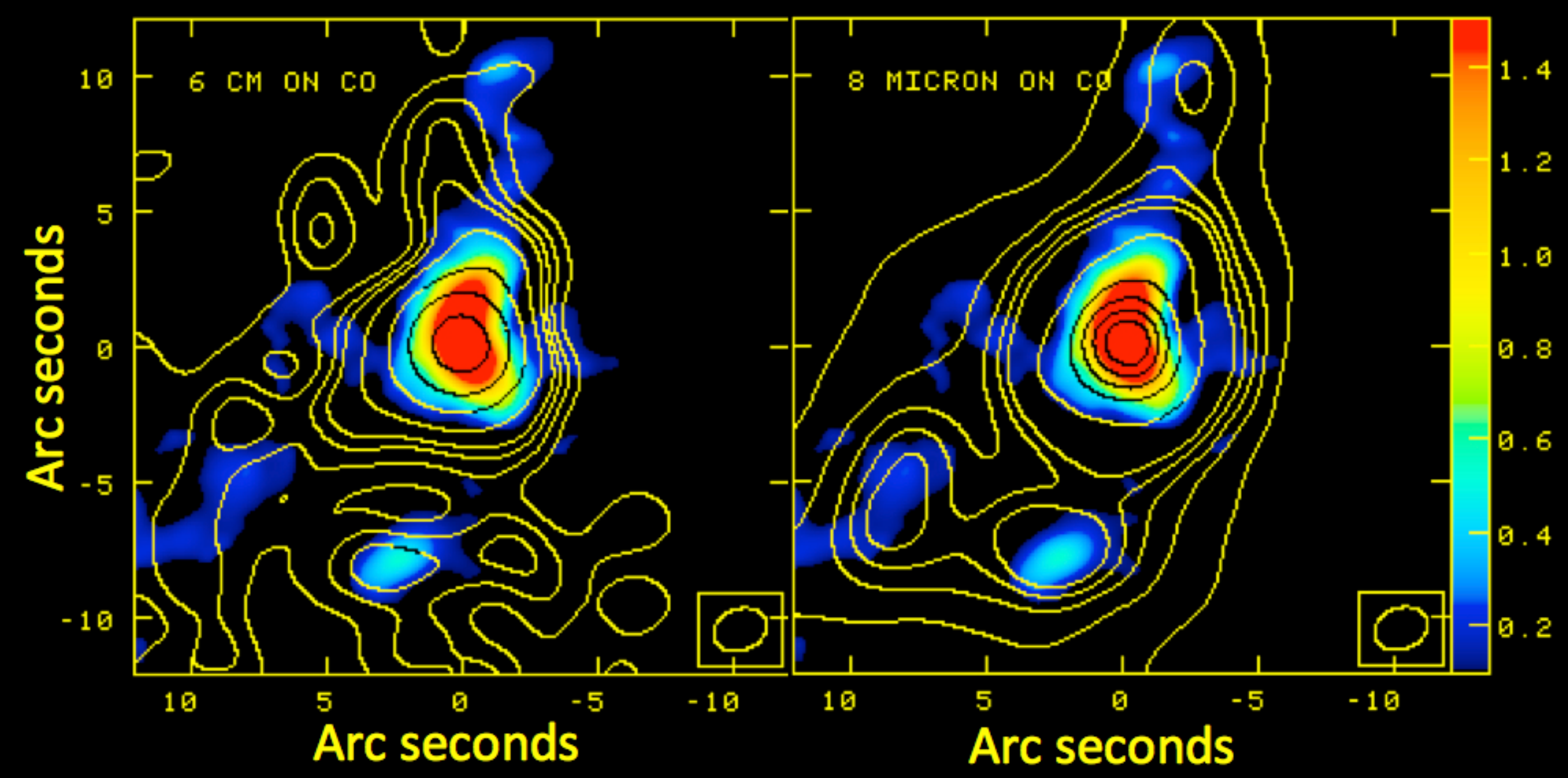}
\caption{Feature {\it i} showing (left) $\lambda 6$ cm radio continuum contours on CO color, with
contour levels of 4, 6, 8, 10, 16, 32, 64, 128 x  the rms noise of 0.016 mJy beam$^{-1}$ (equivalent
to $T_b$ = 0.13 K); and (right) 8 $\mu m$ contours on CO color, with contour levels of
2, 4, 6, 8, 10, 20, 40, 60, 80, 100 MJy sr$^{-1}$. The color wedge is in units of  (Jy beam$^{-1}$)(km s$^{-1}$),
where 1 (Jy beam$^{-1}$)(km s$^{-1}$) corresponds to 88.6 M$_\odot$ pc$^{-2}$.}
\label{featurei68}\end{figure}

Feature {\it i} is an energetic mini-starburst region in the northwestern tip of
NGC 2207; its H$\alpha$, 8$\mu m$, 24$\mu m$, 70$\mu m$, and radio continuum
emission are stronger than anywhere else in the two galaxies, as noted in
\cite{elmegreen00} and \cite{kaufman12}. Figure \ref{feati} highlights Feature {\it
i}. The left image is the color composite  from F439W in blue, F555W in green, and
F814W in red, and the right image is the HST color composite with the CO contours
overlaid in green. Note the biconical C-shaped structure of the CO cloud centered
on Feature {\it i}. Figure \ref{featurei68} shows radio continuum (left) and 8$\mu
m$ (right) contours on CO intensity maps. The 6 cm and 8$\mu m$ emission both peak
at the peak CO emission.

Feature {\it i} contains $8 \times 10^7$ M$_\odot$ of molecular gas in the whole
triangle outlined by stars (as measured with a $14''$ diameter aperture in Paper I)
and $5.9 \times 10^7$  M$_\odot$ in the C-shaped structure.   The southwestern lobe
of the C-shaped structure coincides with the optically opaque conically-shaped dust
cloud (noted in our previous papers) which, in CO,  extends $3.5''$ with its axis
at position angle P.A. $\sim 220^\circ$ from the central star complex. The other
lobe of the C-shaped structure is along  the spiral arm.


The central star complex, located at the peak CO column density, has an SED-fitted
log mass in $M_\odot$ of $6.31\pm0.09$ and a log age
of $5.8\pm0.4$ (0.63 Myr). SExtractor found 13 other star-forming regions within the bright
triangle shape in Figure \ref{feati}, ranging in diameter from 60 pc (12 pixel
area) to 100 pc (42 pixel area). The average log mass of these thirteen is
$5.2\pm0.4$ (in $M_\odot$) and the average log age is $6.9\pm0.8$ (7.9 Myr).
\cite{smith14} derived an age of 5.7 Myr for a 10'' diameter aperture around
Feature {\it i} from the H$\alpha$ equivalent width. The ratio of the central star
complex mass to the CO mass in the C-shaped contour structure of Figure \ref{feati}
is 3.5\%.

Paper I notes that Feature {\it i} (measured with a $14''$ diameter aperture) has an SFR
of $1.6 M_\odot$ yr$^{-1}$, very high compared with its gas content, and the gas is
HI-dominated in the $14''$ aperture. The total neutral hydrogen mass in this aperture is $1.9\times 10^8$ $M_\odot$.
Thus for star complexes with ages less than 30 Myr in Feature {\it i}, the ratio of the summed star complex mass to total neutral hydrogen mass is 1\%.


\subsection{NGC 2207 Nuclear Region}
\begin{figure}\epsscale{1.15} \plotone{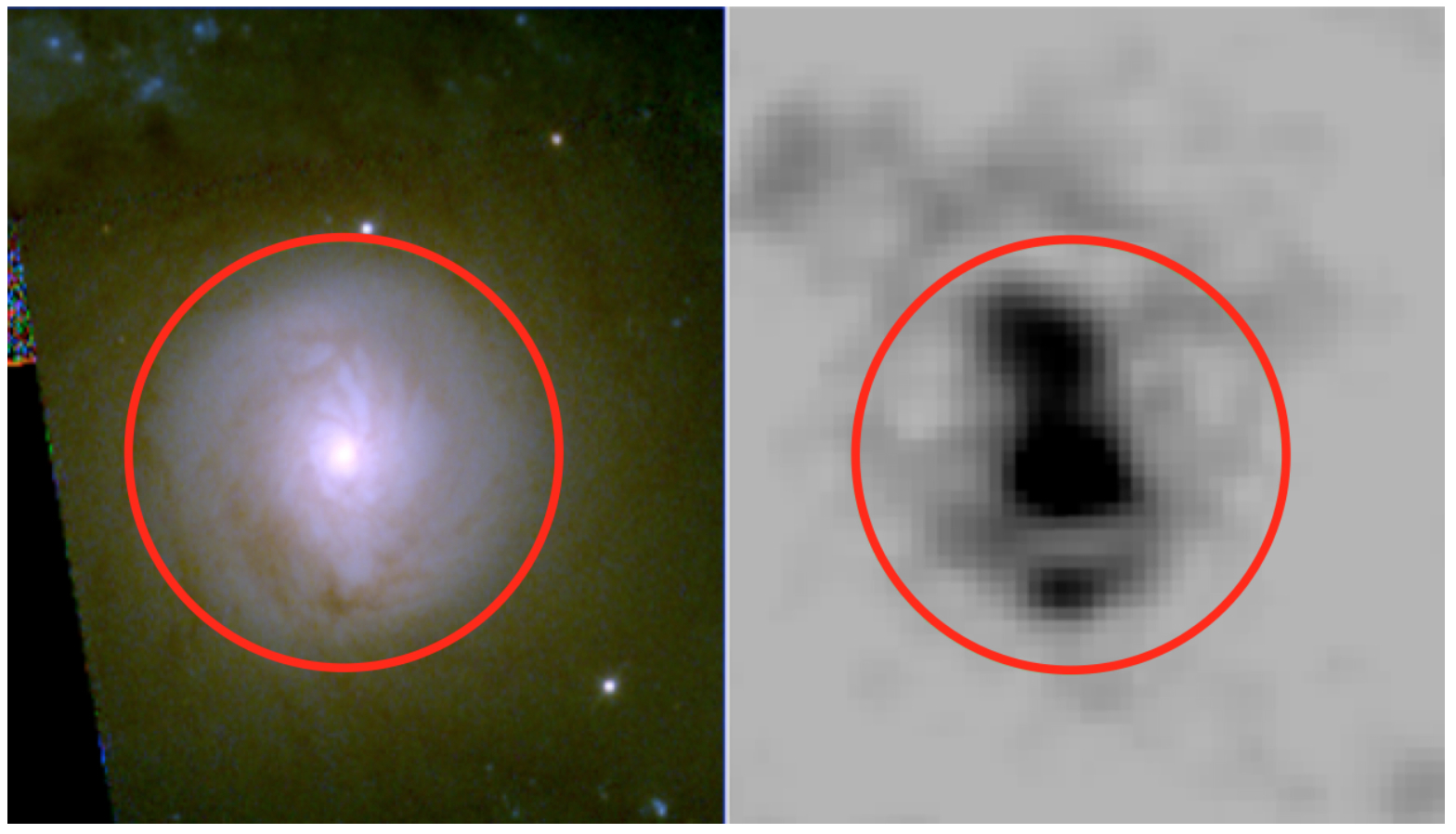}
\caption{(left) Nuclear region of NGC 2207 shown in HST WFPC2 F439W, F555W, F814W filters; (right) CO on the same scale as  the optical color image. The red circle is a fiducial marker for the two images.} \label{nuc}\end{figure}

NGC 2207 has bright CO in the nuclear region, as shown in Figure \ref{nuc}. The
left-hand image shows HST WFPC2 F439W in blue, F555 in green, and F814W in red in a
color composite. The central region of NGC 2207 has complex dust lanes with a power
law power spectrum, presumably resulting from turbulence \citep{elm98,mont99}. The
right-hand figure shows CO intensity in grayscale. The red circle is a fiducial
reference for both images, which are on the same scale.

\begin{figure}\epsscale{1.15} \plotone{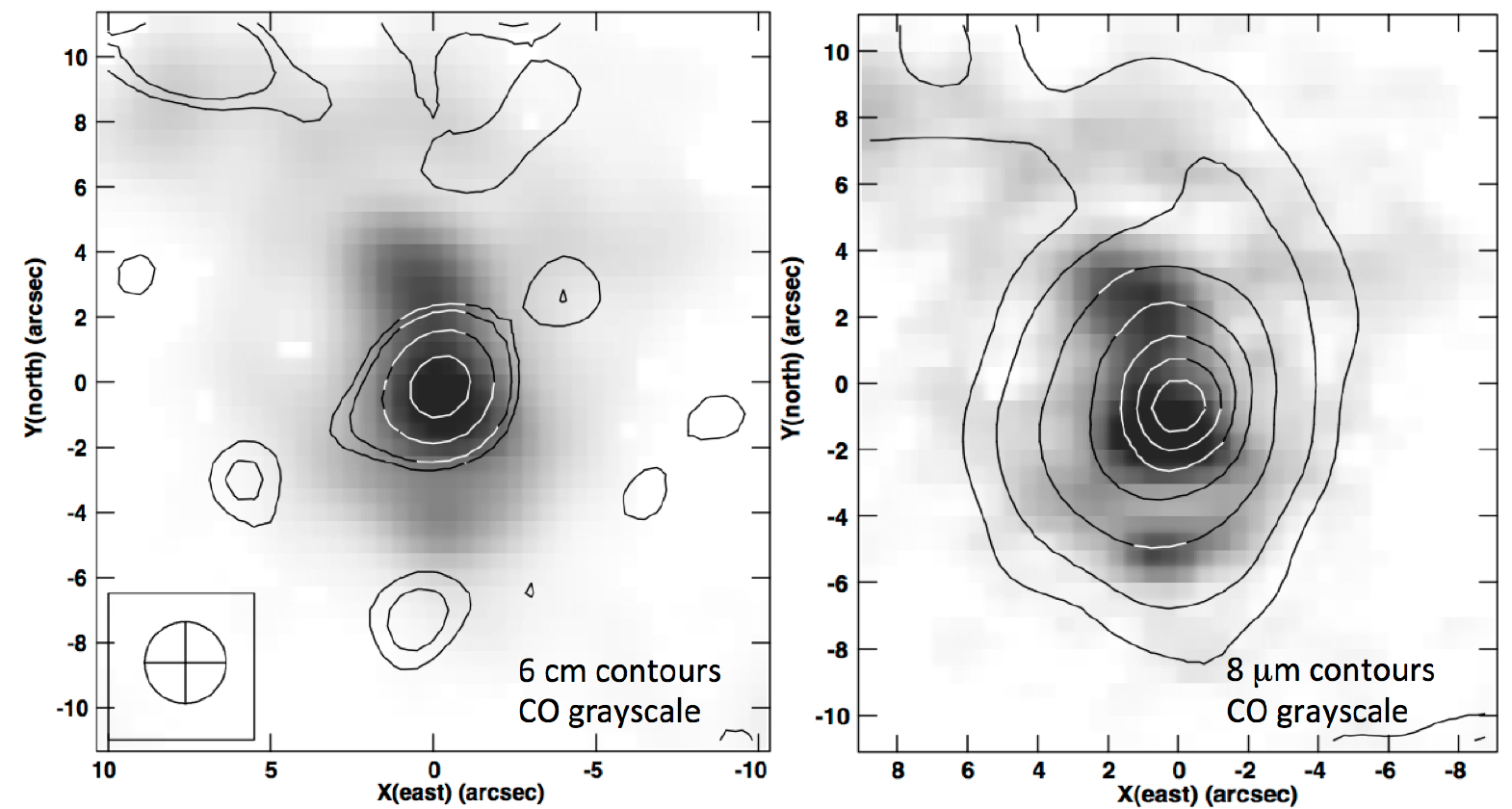}
\caption{(left) Overlay of 6 cm radio continuum contours on grayscale CO emission for the nuclear region of NGC 2207; the contours are at 3, 4, 8, 16, 32, 64 times the rms noise of 0.016 mJy beam$^{-1}$.  (right) Overlay of non-stellar 8$\mu m$ contours on CO grayscale intensity. The contour levels are at
3, 4, 6, 8, 10, 12, 14 MJy sr$^{-1}$. } \label{nuc-cont}\end{figure}

There is a double CO lobe slightly offset from the center of NGC 2207; the northern
lobe coincides with optical dust lanes but does not show up at other wavelengths.
South of the double lobe is another CO cloud that coincides with optical dust features.
The ratio of the northern to southern lobe peak intensity is 0.71. The approximate
size of each CO lobe is 4$^{\prime\prime}$ $\times$ 3$^{\prime\prime}$,
corresponding to 680 $\times$ 510 pc.

A CO ring encircles the central region of NGC 2207, offset slightly north of
center, which also shows up in the combined images of both galaxies in Figure
\ref{I2163B8CO}. The CO ring does not have a counterpart in other wavelengths.
Figure \ref{nuc-cont} shows an enlargement of the central region, with 6 cm
contours on CO grayscale on the left (with the CO image convolved to the 2.5''
resolution of the radio image), and non-stellar 8 $\mu m$ contours on CO intensity
grayscale on the right.  The non-stellar 8 $\mu m$ component was calculated from
IRAC 4 - 0.232 IRAC 1 using the recipe of \cite{helou} (This overestimates the
correction since IRAC 1 contains some PAH bands.) The non-stellar 8$\mu m$ here is
presumably from excitation by the general radiation field in the nucleus, not
star-forming regions.

The CO emission is elongated, possibly a mini-bar, whereas the radio continuum
emission shows a slight elongation perpendicular to the CO mini-bar and along the
direction of the kinematic major axis. It is possible that the CO ring is a
resonance related to the mini-bar.

The locations of maximum brightness in the N2207 nucleus in 6 cm radio continuum
and 8 $\mu m$
coincide, within the uncertainties, with the Chandra position listed by
\cite{mineo} of a low luminosity AGN X-ray source in N2207, which has a
non-symmetric shape of the central AGN in X-rays. The radio continuum is probably
from the AGN. The molecular gas is probably dissociated at the location of the AGN,
which helps produce the double-lobed appearance. The location of peak CO emission
is 1'' south of the AGN.

\section{Conclusions}

Over 200 large-scale molecular emission regions with masses down to $\sim10^6$
$M_\odot$ have been identified in the interacting galaxies IC 2163 and NGC 2207. These
clouds contain about half of the total CO mass in the two galaxies detected by the
interferometer. The eyelid regions of IC 2163 contain the clouds with the highest
average mass;  all have $\log_{10}(M/M_\odot)$ greater than 6.6, and average 7.2. The ages of the
star complexes in the eyelids are younger than $\sim2\times10^8$ yr, similar to the
time of interaction of the two galaxies, whereas some star complexes in NGC 2207 are
older.  This suggests that processes that formed the eyelids destroyed the complexes
there.

The mass distribution functions of the CO clouds follow a power law distribution, as do
the star complexes identified down to masses of about 10$^5$ $M_\odot$ both in the
eyelid regions and elsewhere in the galaxies. The clouds in NGC 2207 have a steeper
mass distribution function than those in the IC 2163 eyelids or in the inner parts of
the Milky Way, but comparable that those in the outer parts of the Milky Way and M33.
To explain these variations, we considered the possibility that the CO cloud mass
function is steeper than the total cloud mass function in regions with low molecular
fraction as a result of a decrease in the CO mass fraction with increasing total cloud
mass.

The ratio of the summed star complex mass within 500 pc of a molecular cloud
to the cloud mass is $\sim12$\% in these galaxies; within 300 pc, it is $\sim7$\%.
For 40 star-forming regions 2.4 kpc in diameter, the ratio of the summed star complex
mass to the summed $H_2$ mass is $\sim17$\%. These ratios are higher than the
efficiencies of star formation in any one cloud. For complexes younger than 30 Myr
and within 300 pc of a cloud, the ratio of stellar to gas mass is $\sim4$\%, and
for complexes younger than 30 Myr in the 2.4 kpc regions, it is also $\sim4$\%. For
the mini-starburst Feature {\it i}, the ratio of the mass of the central star
complex to the total CO mass around it is 3.5\%.

The average star complex extinctions scale with increasing gas column density,
suggesting that the CO gas is in the midplane and not preferentially on one side
from the interaction of the two galaxies.

The nuclear region of NGC 2207 contains a double-lobed CO cloud, possibly a
mini-bar. It is surrounded by a molecular ring, which could be the result of a
resonance associated with the mini-bar. It contains the most massive cloud in the system.

\vspace{0.3cm}

EB acknowledges support from the UK Science and Technology Facilities Council
[grant number ST/M001008/1]. This paper used the following ALMA data:
ADS/JAO.ALMA\#2012.1.00357.S. ALMA is a partnership of ESO (representing its member
states), NSF (USA) and NINS (Japan), together with NRC (Canada) and NSC and ASIAA
(Taiwan) and KASI (Republic of Korea), in cooperation with the Republic of Chile.
The Joint ALMA Observatory is operated by ESO, AUI/NRAO and NAOJ. The National
Radio Astronomy Observatory is a facility of the National Science Foundation
operated under cooperative agreement by Associated Universities, Inc. This research
used the NASA/IPAC Extragalactic Database (NED) which is operated by the Jet
Propulsion Laboratory, California Institute of Technology, under contract with the
National Aeronautics and Space Administration. We thank the referee, Dr. Erik Rosolowsky, for helpful
comments that improved the clarity of the paper.

\end{document}